\documentclass{llncs}

\usepackage[english]{babel}
\usepackage[utf8]{inputenc}
\usepackage{amsmath,amssymb}
\usepackage{graphicx,tikz,wrapfig,url}
\usepackage[all]{xy}
\xyoption{all}
\usepackage[colorinlistoftodos]{todonotes}
\usepackage[lined]{algorithm2e}
\usepackage{listings}

\title{Querying Geometric Figures Using\\ a Controlled Language,
  Ontological Graphs\\ and Dependency Lattices\thanks{The final publication is available at \texttt{http://link.springer.com}.}}

\author{Yannis Haralambous\inst{1} \and Pedro Quaresma\inst{2}}
\institute{Institut Mines-Télécom, Télécom Bretagne
Computer Science Department
UMR~CNRS~6285 Lab-STICC
Technopôle Brest Iroise
CS 83818, 29238~Brest~Cedex~3, France\and CISUC/Departament of Mathematics, University of Coimbra\\
P-3001-454 Coimbra, Portugal}

\date{\today}

\def\ca#1#2{\begin{tabular}{@{}c@{}}#1\\#2\end{tabular}}

\begin{document}
\maketitle
\begin{abstract}

Dynamic geometry systems (DGS) have become basic tools in many areas of geometry as, for example, in education. Geometry Automated Theorem Provers (GATP) are an active area of research and are considered as being basic tools in future enhanced educational software as well as in a next generation of mechanized mathematics assistants. Recently emerged Web repositories of geometric knowledge, like TGTP and Intergeo, are an attempt to make the already vast data set of geometric knowledge widely available. Considering the large amount of geometric information already available, we face the need of a query mechanism for descriptions of geometric constructions.

In this paper we discuss two approaches for describing geometric figures (declarative and procedural), and present algorithms for querying geometric figures in declaratively and procedurally described corpora, by using a DGS or a dedicated controlled natural language for queries.
\end{abstract}

\section*{Introduction}

Dynamic geometry systems (DGS) distinguish themselves from drawing programs in two major ways. The first is their knowledge of geometry: from a initial set of objects drawn freely in the Cartesian plane (or maybe, on another model of geometry), one can specify/construct a given geometric figure using relations between the objects, e.g., the intersection of two non-parallel lines, a line perpendicular to a given line and containing a given point, etc. Another major feature of a DGS is its capability to introduce dynamics to a given geometric construction moving a (free) basic object always preserving the geometric properties of the construction~\cite{wikipediaLIGS}. 

That is, one uses a DGS by constructing a geometric figure with geometric objects and geometric relations between them, and not by placing points on specific Cartesian coordinates. Most (if not all) DGS possess a formal language for the specification of geometric constructions. In some systems this formal language is explicit, in others it is hidden from the user by the graphical interface. The Intergeo project designed a common format, called I2G, for this formal language which is already accepted by some DGS~\cite{Santiago2010,i2gImplementationTable}.
Geometry automated theorem provers (GATP), being formal systems,
need a formal language to describe geometric conjectures. GATPs are
nowadays mature tools capable of proving hundreds of geometric conjectures~\cite{Chou2001,Jiang2012}.  The I2GATP formal language is an  extension of the formal language used by the DGS. The I2GATP project goal is to define a common language, an extension of the I2G language, to the DGS/GATP tools~\cite{Quaresma2012}.

The design of common languages, and the emergence of Web repositories of geometric knowledge is an attempt to make the
 already vast data set of geometric knowledge widely available. The Intergeo project~\cite{Kortenkamp2009}, GeoThms~\cite{Janicic06a} and TGTP~\cite{Quaresma2011} systems already meet some of these goals, having provided a large data set of geometric information widely available. In these systems the question of querying the geometric construction is not solved, that is, it is not yet possible to query the data set for a construction similar to some other construction, or to query for all constructions having some common geometric properties.
The goal of our research is to develop a search mechanism for geometric constructions (done by a DGS or a GATP) using the different ways of geometric construction descriptions.

%Our approach is \todo{Describe the possible solutions and explain why we chose a given solution.}

\section{What You See and How to Get It: Declarative vs.~Procedural vs. Analytic Figure Description}

%\dedicatory{Ceci n'est pas une pipe\\\textup{René Magritte}}
\begin{wrapfigure}{r}{0pt}
\includegraphics{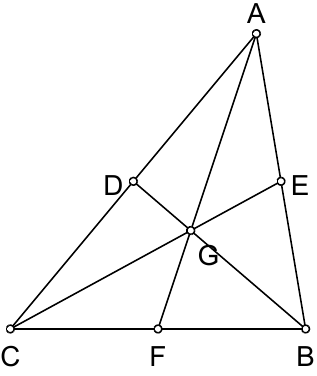}
\caption{Construction 13 of the corpus.\label{013}}
\end{wrapfigure}
On Fig.~\ref{013} the reader can see (a visual representation of) the centroid theorem, a simple geometric figure taken from the TGTP corpus~\cite[Fig.~13]{Quaresma2011}.

There are many approaches for describing this figure:
\begin{itemize}
\item the \emph{declarative} one: “we have points $A,E,B$ on the same line such that $AE=EB$, points $A,D,C$ on the same line such that $AD=DC$, points $C,F,B$ on the same line such that $CF=FB$, and displayed are $AC, BC, AB, AF, BD, CE$. The intersection of $BD$, $CE$, $AF$ is called~$G$”;
\item the \emph{procedural} one: “draw segments $AB$, $AC$, $BC$, take their midpoints $E$, $D$, $F$, draw $AF$, $BD$ and $CE$, take the intersection of $BD$ and $CE$ and call it~$G$”;
\item the \emph{analytic} one: “points $A$, $B$, $C$ have coordinates $(35,40)$, $(10,10)$, $(40,10)$; points $D$, $E$, $F$, have coordinates $(37.5,25)$, $(22.5,25)$, $(25,10)$; segments have coordinates $((35,40),(40,10))$, $((35,40),(10,10))$”, etc.
\end{itemize}

In this paper we will concentrate on the \emph{procedural} and \emph{declarative} descriptions of a figure. 

The declarative description is about ``what the parts of the figure are and how they relate to each other,'' while the procedural description is about ``how to construct the figure.'' In the former we can supply arbitrary (and potentially redundant) information about the figure; in the latter we provide only instructions that result into the given figure.

The first problem we encountered when querying figures was the fact that a given figure can often be constructed (and hence, procedurally described) in several ways. For example, Fig.~\ref{013} can be procedurally obtained in (at least) the following two ways (cf.~\cite{diffeor} for the second):

\medskip

\noindent\begin{tabular}{@{}p{.5\textwidth}p{.5\textwidth}@{}}
start with points $A$, $B$, $C$\hfill\break
draw midpoints of $AB$, $AC$, $BC$\hfill\break
call them $D$, $E$, $F$\hfill\break
draw the segments&start with points $D$, $E$, $F$\hfill\break 
draw a line at $F$ parallel to $DE$\hfill\break
draw a line at $E$ parallel to $DF$\hfill\break
draw a line at $D$ parallel to $EF$\hfill\break
call the intersections $A$, $B$ and $C$\hfill\break
remove the lines\hfill\break
draw the segments
\end{tabular}

\medskip

That is, we can start with the triangle and find the midpoints, or we can start with the midpoints and find the triangle.

\setcounter{footnote}{0}
Both the DGS to be used (GeoGebra \cite{geogebra}) and the controlled language we will define (\S\,\ref{cnl}) are \emph{procedural}, and hence describe a figure by its construction. As there are many constructions resulting in the same figure, we concluded that our search system should better use a \emph{declarative} approach. For this (see also~\cite{H-Q2012}), we convert procedural descriptions into declarative ones and represent the information they contain by the use of \emph{ontological graphs}. This operation is done both for the search corpus and for the queries, so that a figure query becomes the search of a graph pattern inside a corpus of ontological graphs.

The second problem we encountered is that procedural descriptions are sometimes lacunary, provided the correct visual result is obtained. For example, in the procedural description of Fig.~\ref{013}, as it is included in our corpus, the creator of the figure has defined $G$ as being the intersection of $BD$ and $CE$, without going any further. Since the goal was to obtain the correct visual representation of the figure, it was not necessary to state that $G$ is also the intersection of $AF$ and $BD$ as well as of $AF$ and $CE$. This means that, after conversion into the declarative representation, the information provided in it will lack these facts. 

Inference can fill some of the gaps and make a declarative description more complete. For example it can detect parallelisms or orthogonality relations that are not explicitly stated.\footnote{In a future development we plan to use the deductive database method to find all the fix-points for a given construction, finding in this way the missing facts~\cite{Chou2000}.} 

The way we propose to solve this problem is by going the ``other way around'': instead of making the corpus richer, we can weaken the query. This method is called \emph{query reduction} and is useful when the query contains too much information and cannot be found in the corpus. 

The problem then is, how do we reduce the query? Indeed, when in front of an ontological graph query where all ingredients of the query figure have become nodes, and their relations have become edges, how do we choose the most suitable nodes or edges to remove?

It is the procedural description of the query that provides us with an answer\footnote{Well understood, the answer is not unique since it strongly depends on the way the figure has been constructed, which is not unique.} to this question. From the procedural data, we build a \emph{dependency lattice} of the query figure. The lattice structure provides us with the nodes to remove, and the order in which to remove them.

For these reasons, we have developed, and will discuss in this paper, both procedural and declarative descriptions of geometric figures. Thanks to their complementarity we obtain en efficient geometric figure search system.

\section{Ontological Graphs}

\subsection{Describing a Geometric Figure by an Ontological Graph}\label{ontographs}

In the following we will use an ontology specific to geometric figures on the plane. This ontology contains concepts:
\begin{itemize}
\item \emph{point}: a point of the plane;
\item \emph{segment}: a segment, defined by two points. It has an attribute ``length'' which induces a relation of ``ratio'' among segment instances;
\item \emph{line}: a line, defined by two points or in some other way (for example, by a point and a property like perpendicularity);
\item \emph{conic}: a conic defined in various ways, and, in particular, a circle, defined by its center point and another point;
\item \emph{angle}: the angle of two segments/lines, it has an attribute \emph{value} which can have a numeric value or the modal value ``straight''.
\end{itemize}

The relations between instances will be\footnote{The list is not exhaustive.}:
\begin{itemize}
\item \emph{belongs\_to}: a relation whose domains are both points (belonging to segments, lines, circles and angles), and segments (belonging to lines);
\item \emph{has\_ratio}: can be used for lengths and angle values. It is a 3-ary reified relation, the members of which are the nominator, denominator and ratio value;
\item \emph{is\_center\_of}: connects a point with the circle of which it is the center;
\item \emph{is\_parallel\_to}: connects two parallel lines (using inference, we will find all parallel lines);
\item \emph{is\_perpendicular\_to}: connects two perpendicular lines or segments (using inference, we will find all perpendicular lines or segments, knowing that the perpendicular of a perpendicular is a parallel);
\item \emph{is\_radius\_of}: connects a segment with the circle of which it is the radius.
\end{itemize}

These concepts and relations have been inspired by the element types of DTD \textsf{GeoCons.dtd}~\cite{Quaresma08} (the ontology does not cover XML elements \textsf{towards}, \textsf{trans\-lation}, \textsf{rotation} which are useful for drawing but do not affect the ontological graph of the figure) and of GeoGebra XML schema \textsf{ggb.xsd}~\cite{geogebra}.

Every figure becomes a graph of instances of concepts and of relations. Not only this approach is independent of the way the figure has been constructed, but it is also independent of instance names and allows to focus on the network of relations between the ingredients of the figure.

Our choice of concepts and relations makes some graphical constructions obtainable by a single relation, for example: ``$AB\perp BC$''
\begin{center}
\includegraphics{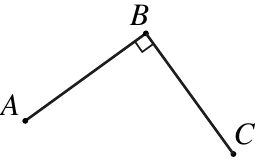}
\end{center}
is represented by instances $A,B,C$ (points), $AB$ and $BC$ (segments) and $\angle ABC$ (angle), and the following graph of relations:

\begin{center}
\begin{tikzpicture}[->,x=1.5cm,y=.9cm,auto,node distance=3cm,
  thick,main node/.style={font=\footnotesize}]

\node[main node] (1) at (0,1)  [fill=green!40] {$A$};
\node[main node] (2) at (1,1)  [fill=red!20] {$AB$};
\node[main node] (3) at (2,1)  [fill=green!40] {$B$};
\node[main node] (4) at (3,1)  [fill=red!20] {$BC$};
\node[main node] (5) at (4,1)  [fill=green!40] {$C$};
\node[main node] (6) at (2,0) [fill=blue!20] {\ca{$\angle ABC$}{value=$\perp$}};

\path[every node/.style={font=\sffamily\small}]
(1) edge (2)
(1) edge (6)
(2) edge (6)
(3) edge (2)
(3) edge (6)
(3) edge (4)
(4) edge (6)
(5) edge (4)
(5) edge (6)
(3) edge (4);
\end{tikzpicture}
\end{center}

\noindent where solid arrows denote the \emph{belongs\_to} relation, and $\perp$ is the ``right angle'' value of the value attribute.

Other constructions, although trivial, are more difficult to encode. For example: ``$AB$ is tangent at circle $c$ at point $B$''
\begin{center}
\includegraphics{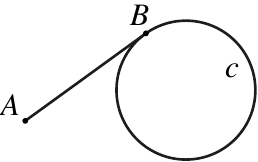}
\end{center}
cannot be encoded by a single relation. We need to use the radius $BO$ and say that $B\in c\wedge AB\perp BO$
\begin{center}
\includegraphics{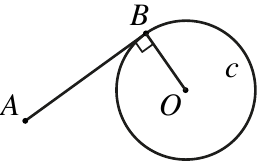}
\end{center}
and the graph of relations will be

\begin{center}
\begin{tikzpicture}[->,x=1.5cm,y=.9cm,auto,node distance=3cm,
  thick,main node/.style={font=\footnotesize}]

\node[main node] (1) at (0,1)  [fill=green!40] {$A$};
\node[main node] (2) at (1,1)  [fill=red!20] {$AB$};
\node[main node] (3) at (2,1)  [fill=green!40] {$B$};
\node[main node] (4) at (3,1)  [fill=red!20] {$BO$};
\node[main node] (5) at (4,1)  [fill=green!40] {$O$};
\node[main node] (6) at (2,0) [fill=blue!20] {\ca{$\angle ABO$}{value=$\perp$}};
\node[main node] (8) at (4,0) [fill=black!20] {$c$};

\path[every node/.style={font=\sffamily\small}]
(1) edge (2)
(1) edge (6)
(2) edge (6)
(3) edge (2)
(3) edge (6)
(3) edge (4)
(4) edge (6)
(5) edge (4)
(5) edge (6)
(3) edge (4)
(4) edge [dotted] (8)
(5) edge [dashed] (8);
\end{tikzpicture}
\end{center}
\noindent where the dotted arrow represents the relation \emph{is\_center\_of} and the dashed one, the relation \emph{is\_radius\_of}.

The ontological graph of a geometrical figure can rapidly increase in size. Its generation is done in a two-step process:
\begin{enumerate}
\item every XML element of the figure description is converted into a set of concepts and relations;
\item then, inference is applied to generate additional relations:
\begin{enumerate}
\item we calculate the transitive closure of parallelism and orthogonality relations ($a \mathrel{/\!/} b\wedge b\mathrel{/\!/} c\vdash a\mathrel{/\!/}c$ and $a\perp b\wedge b\perp c\vdash a\mathrel{/\!/} c$);
\item nodes having equal lengths or equal angle values by construction\footnote{We emphasize the fact that equality is explicitly given in the construction and is not the result of measurements between objects of the figure.} obtain a \emph{has\_ratio} relation with value attribute equal to~$1$;
\item if necessary, angles are instantiated for every pair of segments with a common point.
\end{enumerate}
\end{enumerate}

Our corpus of 134 figures, encoded as an XML file of 3,137 elements resulted into graphs of a total of 5,282 concept instances and 10,211 relation instances.

\subsection{Example}

Take Fig.~\ref{013} representing Figure~13 of the corpus (illustrating the fact that medians intersect at the barycenter of the triangle).
The construction, as given in the XML file, takes arbitrary points $A,B,C$, defines $D$ (resp. $E$, $F$) as the midpoint of $AC$ (resp. $AB$, $BC$), and $G$ as the intersection of $BD$ and $CE$. Furthermore, the segment $AF$ is drawn.

The ontological graph will contain concepts for points $A,B,C,D,E,F,G$, and segments $AB,AC,AD,AE,AF,BC,BD,BE,BF,CD,CE,CF$. The relations will all be of type \emph{belongs\_to}, except for some 3-ary \emph{has\_ratio} relations representing equal lengths. In Fig.~\ref{o013}, unlabelled arrows denote the \emph{belongs\_to} relation.

\begin{figure}[ht]
\begin{center}
\begin{tikzpicture}[->,x=1.5cm,y=.9cm,auto,node distance=3cm,
  thick,main node/.style={font=\footnotesize}]

\node[main node] (3) at (1,4)  [fill=green!40] {$A$};
\node[main node] (4) at (2,4)  [fill=red!20] {$AE$};
\node[main node] (5) at (4,5)  [fill=green!40] {$E$};
\node[main node] (6) at (6,4) [fill=red!20] {$BE$};
\node[main node] (7) at (7,4) [fill=green!40] {$B$};
\node[main node] (8) at (4,4) [fill=red!20] {$AB$};
\node[main node] (10) at (1,2.5)  [fill=red!20] {$AD$};
\node[main node] (11) at (2,2.5)  [fill=red!20] {$AC$};
\node[main node] (12) at (3,2.5)  [fill=red!20] {$CE$};
\node[main node] (13) at (4,2.5) [fill=green!40] {$G$};
\node[main node] (14) at (5,2.5)  [fill=red!20] {$BD$};
\node[main node] (15) at (6,2.5)  [fill=red!20] {$BC$};
\node[main node] (16) at (7,2.5)  [fill=red!20] {$BF$};
\node[main node] (18) at (1,1)  [fill=green!40] {$D$};
\node[main node] (19) at (2,1)  [fill=red!20] {$CD$};
\node[main node] (20) at (4,0)  [fill=green!40] {$C$};
\node[main node] (21) at (6,1) [fill=red!20] {$CF$};
\node[main node] (22) at (7,1) [fill=green!40] {$F$};
\node[main node] (25) at (4,1) [fill=red!20] {$AF$};
\node[main node] (26) at (4,6) [fill=yellow!20] {\emph{has\_ratio}};
\node[main node] (27) at (5.5,6) {1};
\node[main node] (28) at (0.5,0) [fill=yellow!20] {\emph{has\_ratio}};
\node[main node] (29) at (2,0) {1};
\node[main node] (30) at (7.5,0) [fill=yellow!20] {\emph{has\_ratio}};
\node[main node] (31) at (6,0) {1};

\path[every node/.style={font=\sffamily\small}]
(3) edge (4)
(3) edge [bend right,looseness=0.5] (8)
(3) edge (10)
(3) edge (11)
(3) edge (25)
(5) edge (4)
(5) edge (12)
(5) edge (6)
(5) edge (8)
(7) edge (6)
(7) edge [bend right,looseness=0.5] (8)
(7) edge (14)
(7) edge (15)
(7) edge (16)
(13) edge (12)
(13) edge (14)
(18) edge (10)
(18) edge (11)
(18) edge (14)
(18) edge (19)
(20) edge (11)
(20) edge (12)
(20) edge (15)
(20) edge (19)
(20) edge (21)
(22) edge (15)
(22) edge (16)
(22) edge (21)
(22) edge [bend right,looseness=0.5] (25)
(26) edge node [left] {nominator} (4)
(26) edge node [right] {denomin.} (6)
(26) edge node {value} (27)
(28) edge node [left] {nominator} (10)
(28) edge node [above] {denomin.} (19)
(28) edge node {value} (29)
(30) edge node [right] {nominator} (16)
(30) edge node [below] {denomin.} (21)
(30) edge node {value} (31)
;
\end{tikzpicture}
\end{center}
\caption{The ontological graph of Fig.~\ref{013}.\label{o013}}
\end{figure}

\subsection{Querying Ontological Graphs}

To be able to search in a corpus, we convert all figures of the corpus into ontological graphs and we store them in a graph database (we use a neo4j database~\cite{Cypher}). The user query is a figure drawn by using a DGS, or a query using the controlled query language (\S\ref{cnl}). This figure or CQL statement is converted into an ontological graph on-the-fly, and then into a Cypher query (Cypher is the query language of the neo4j graph database system). The query is send to the database and returns graph instances containing the query as sub-graph.

At this step, ranking is performed to present the results to the user in a pertinent way. Our ranking criterion (which we will try to improve in the future) is the ratio between number of nodes and relations of the query and the number of nodes and relations of the matched graphs. Using this criterion we obtain first the smallest figure possible figure containing the query subgraph. We intend to use graphical mechanisms to highlight the matched pattern in the resulting graphs by using, for example, a different color.

%\subsection{Querying dependency lattices}

%The principle is the same as for ontological graphs: the user draws a figure in the GeoGebra applet or uses the controlled language (cf.~\ref{cnl}), this figure is converted into a dependency lattice, and this lattice is used as a search pattern in the graph database of the corpus.

%As we will see in \S~\ref{incompletedl}, the lattice structure strongly constrains the set of nodes that can be removed to obtain reduced queries. Contrarily to ontological graphs, to obtain a reduced query, one can only remove specific nodes, namely those that are parents of the sink node, their parents, and so forth. Indeed, if we remove, for example, node $B$ of Fig.~\ref{d013} then no descendant of $B$ can be calculated anymore, and the only remaining nodes are $A$, $C$ and $D$.

\section{The Controlled Query Language}\label{cnl}

In some cases the user may not wish to use the DGS to build the query, either because it is cumbersome to use or because it does not provide the necessary abstractions. We propose, as an alternative to the DGS, a \emph{controlled query language} that allows the (procedural) description of a figure in a way that is simple and close to natural language.

\subsection{Description of the controlled query language}

Here is the grammar of the controlled query language:
\begin{verbatim}
S -> query
query -> sents drawvp PERIOD
query -> sents PERIOD
drawvp -> DRAW ents
sents -> sent SEMICOLON sents
sents -> sent
sent -> nps vrb pps
sent -> nps vrb
ent -> INST LABEL
ent -> LABEL
vrb -> VERB ADJE
vrb -> VERB NOUN
vrb -> VERB
pps -> ents pents
pps -> pents
pps -> ents
pents -> pent pents
pents -> pent
pent -> PREP ents
ents -> ent AND ents
ents -> ent
nps -> ents
NOUN -> /(midpoint|foot|mediatrix|intersection|bisector)[s]?/
INST -> /(points|segments|lines|angles|circles|centers|point|
         segment|line|angle|circle|center)/
VERB -> /(is|are|intersect[s]?|connect[s]?)/
ADJE -> /(perpendicular|parallel|defined|right)/
SEMICOLON -> /;/
AND -> /(,|and)/
DRAW -> /draw/
PREP -> /(at|of|by|to|on)/
PERIOD -> /\./
LABEL -> /[A-Z]([_]?[0-9]+)?(-[A-Z]([_]?[0-9]+)?)*/
\end{verbatim}
and here are its rules\footnote{We will use \textsf{sans serif} font for illustrating the controlled language.}:
\begin{enumerate}
\item Every query is of the form:
\begin{quote}
[list of sentences separated by ;] \textsf{draw} [list of instances separated by \rlap{,].}
\end{quote}
\item An \emph{instance} consists of a type and a name (or just a name, if there is no ambiguity). It is written in the form ``\textsf{type} [name]''.
\item A \emph{primitive instance} is an instance of a point, a line or a circle.
\item The name of a primitive instance matches the regular expression\\ \verb=[A-Z]([_]?[0-9]+)?=.
\item Names of non-primitive instances are composite: they are written by joining names of points using the hyphen character (for example: \textsf{segment A-B}).
\item An instance type can be followed by more than one instance, in that case it is written in plural form and the instances are separated by commas (for example: \textsf{points A, B, C}).
\item The first part of a query defines (and draws) new instances, the second draws already known instances.
\item The following sentences can be used\footnote{All type names are optional except for \textsf{center} in~(h).}:
\begin{enumerate}
\item \textsf{line ? intersects line ? at point ?}
\item \textsf{point ? is the midpoint of segment ?}
\item \textsf{line ? is perpendicular to line ? at point ?}
\item \textsf{point ? is the foot of point ? on line ?}
\item \label{nonsyn}\textsf{line ? is the mediatrix of segment ?}
\item \textsf{line ? is parallel to line ? at point ?}
\item \textsf{line ? connects points ?, ?}
\item \textsf{circle ? is defined by center ? and point ?}
\item \textsf{points ?, ? are the intersections of circles ?, ?}
\item \textsf{points ?, ? are the intersections of circle ? and line ?}
\item \textsf{line ? is the bisector of angle ?}
\item \textsf{angle ? is right}
\end{enumerate}

\medskip

\item All sentences have plural forms where arguments are distributed at all positions and separated by commas (for example: \textsf{lines L\_1, L\_2, L\_3 connect points A\_1, B\_1, A\_2, B\_2, A\_3, B\_3}, which means that $\{A_i,B_i\}\subset L_i$). 
\item In all sentences except (\ref{nonsyn}), the terms \textsf{segment} and \textsf{line} are synonymous, with the syntactic difference that \textsf{segment} must be followed by a composite name (for example \textsf{A-A\_1}), while \textsf{line} must be followed by a primitive name, since ``line'' is a primitive instance. 
\item Some variation is allowed, for example \textsf{and} is a synonym of the comma, articles \textsf{the} in front of nouns are optional.
\item Queries end by a period `\textsf{.}'.
\end{enumerate}

Here is, for example, a description of Fig.~\ref{013} in the controlled query language:
\begin{quote}\label{cqlsample}
\textsf{D, E, F are midpoints of A-C, A-B, B-C ; C-E intersects B-D at G ; draw A-C, A-F, A-B, B-C, B-D, C-E.}
\end{quote}

The query language is compiled, producing a Cypher query, which is then submitted to the graph database exactly as when using the DGS. The compiler has been developed using the Python PLY package \cite{PLY}.

\subsection{Future plans for the controlled language}\label{fpcnl}

In future versions of the controlled language, we plan to introduce the possibility of extending the query ontology by introducing new concepts and/or new relations. For example, it may be interesting to define a type \textsf{square} as
\begin{quote}
\textsf{Points A, B, C, D form a square A-B-C-D when we draw equal segments A-B, B-C, C-D, D-A where angles A-B-C, B-C-D, C-D-A, D-A-B are right.}
\end{quote}

This would allow queries of the form (which will draw the notorious figure of the Pythagorean theorem)
\begin{quote}
\textsf{Angle A-B-C is right ; A-C-C\_1-A\_2, A-A\_1-B\_1-B, C-B-B\_2-C\_2 are squares.}
\end{quote}

\section{Reduced Queries}

The algorithms we describe in this paper can be quite successful in finding exact matches of queries in the corpus. But what happens when the figures in the corpus match only partially the query?

\subsection{Ontological Graphs}

For example, let us consider Fig.~\ref{o013} anew. The ontological graph of the figure has been build solely using the XML data of Figure~13 of the corpus (cf. Fig.~\ref{013}). What is not visible on Fig.~\ref{013} is the fact that $G$ has not been defined as lying on $AF$, and hence the \emph{belongs\_to} edge between $G$ and $AF$ is missing in the ontological graph.

This is also reflected in the CQL query example we gave in \S\,\ref{cqlsample}, where we request that \textsf{C-E intersects B-D at G} but not that \textsf{A-F intersects B-D at G}, probably because this could be inferred from the previous one, if we had the external Euclidean Geometry knowledge of the fact that the three medians of a triangle have a common intersection.

Nevertheless, the user seeking Fig.~\ref{013} is not necessarily aware of this subtlety, and will search for ``a triangle with three medians,'' which will result in an ontological graph similar to the one of Fig.~\ref{o013} but containing an additional edge $G\to AF$, and this graph, of course, will not match Figure~13 of the corpus, since it is not a sub-graph of it.

To solve this problem, as long as a query does not return any results, we retry with \emph{reduced queries}, in the sense of the same query graph with one or more instances (or relations) removed.

But how do we decide which nodes and edges to remove from a query, and in what order? The answer to this question is provided by \emph{dependency lattices}, described in the next section.

\subsection{Dependency Lattices}

Let us return to the \emph{procedural} approach of describing geometric figures. How do we describe a figure using the operations that led to its construction?

Strictly speaking, such a description would require a Berge-acyclic hyper-graph \cite[\S3]{Fagin:1983dy}, where each operation would be a hyper-edge, connecting the input (the set of known nodes) and the output (the set of new nodes), for example, in the case of the \emph{midpoint} operation on segment $AC$, the hyper-edge would connect $\{A,C\}$ (input) and $\{B\}$ (output).

But there is a simpler way. In fact, it  suffices for our needs to represent \emph{dependencies} as edges of a directed graph. For example, in the midpoint example, $B$ is dependent of $A$ and $C$, since the latter have been used to calculate the former:

{\centering
\begin{tikzpicture}[->,x=1.5cm,y=.9cm,auto,node distance=3cm,
  thick,main node/.style={font=\footnotesize}]

\node[main node] (1) at (0,1)  [fill=green!40] {$A$};
\node[main node] (2) at (1,0)  [fill=green!40] {$B$};
\node[main node] (3) at (2,1)  [fill=green!40] {$C$};

\path[every node/.style={font=\sffamily\small}]
(1) edge node [below] {m} (2)
(3) edge node [below] {m} (2)
;
\end{tikzpicture}

}

By adding a ``global source node'' (located above all source nodes) and a ``global sink node'' (underneath all ``final results''), this graph becomes a \emph{lattice}, the partial order of which is the dependency relation.

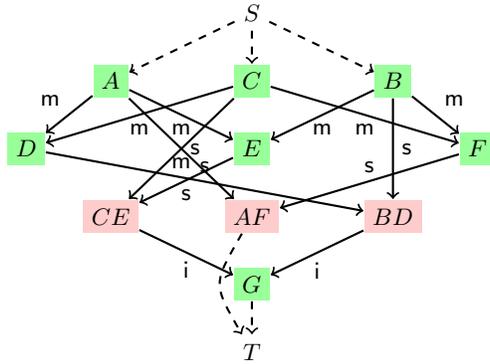
\begin{figure}[ht]
{\centering
\begin{tikzpicture}[->,x=1.5cm,y=.9cm,auto,node distance=3cm,
  thick,main node/.style={font=\footnotesize}]

\node[main node] (1) at (2,4)  [fill=white] {$S$};
\node[main node] (2) at (0.75,3)  [fill=green!40] {$A$};
\node[main node] (3) at (2,3)  [fill=green!40] {$C$};
\node[main node] (4) at (3.25,3)  [fill=green!40] {$B$};
\node[main node] (5) at (0,2)  [fill=green!40] {$D$};
\node[main node] (6) at (2,2) [fill=green!40] {$E$};
\node[main node] (7) at (4,2) [fill=green!40] {$F$};
\node[main node] (8) at (3.25,1) [fill=red!20] {$BD$};
\node[main node] (9) at (0.75,1)  [fill=red!20] {$CE$};
\node[main node] (10) at (2,0)  [fill=green!40] {$G$};
\node[main node] (11) at (2,-1)  [fill=white] {$T$};
\node[main node] (12) at (2,1)  [fill=red!20] {$AF$};

\path[every node/.style={font=\sffamily\small}]
(1) edge [dashed] (2)
(1) edge [dashed] (3)
(1) edge [dashed] (4)
(2) edge node [above left] {m} (5)
(2) edge node [below] {m} (6)
(3) edge node [below] {m} (5)
(3) edge node [below] {m} (7)
(4) edge node [below] {m} (6)
(4) edge node [above right] {m} (7)
(5) edge node [above] {s} (8)
(4) edge node [right] {s} (8)
(6) edge node [below] {s} (9)
(3) edge node [right] {s} (9)
(8) edge node [below] {i} (10)
(9) edge node [below] {i} (10)
(10) edge [dashed] (11)
(7) edge node [above] {s} (12)
(2) edge node [below] {m} (12)
(12) edge [bend right, looseness=1.5, dashed] (11)
;
\end{tikzpicture}

}
\caption{The dependency lattice of Fig.~\ref{013}, as it is procedurally described in the corpus.\label{d013}}
\end{figure}

On Fig.~\ref{d013}, the user can see the dependency lattice of Fig.~\ref{013}. We have used only nodes that are used in calculations, so that, for example, segments $AB$, $AE$, etc. do not appear in the lattice. $S$ and $T$ are the global source and global sink nodes, they are connected to nodes of the lattice by dashed arrows. Full arrows represent operations and are labelled by their initial letters (m = midpoint, s = segment drawing, i = intersection).

\subsection{Using Dependency Lattices for Reduced Queries}

\begin{figure}[ht]
\begin{tabular}{@{}c@{\quad}c@{}}
\begin{tikzpicture}[->,x=1.5cm,y=.9cm,auto,node distance=3cm,
  thick,main node/.style={font=\footnotesize}]

\node[main node] (1) at (2,4)  [fill=white] {$S$};
\node[main node] (2) at (0.75,3)  [fill=green!40] {$A$};
\node[main node] (3) at (2,3)  [fill=green!40] {$C$};
\node[main node] (4) at (3.25,3)  [fill=green!40] {$B$};
\node[main node] (5) at (0,2)  [fill=green!40] {$D$};
\node[main node] (6) at (2,2) [fill=green!40] {$E$};
\node[main node] (7) at (4,2) [fill=green!40] {$F$};
\node[main node] (8) at (3.25,1) [fill=red!20] {$BD$};
\node[main node] (9) at (0.75,1)  [fill=red!20] {$CE$};
\node[main node] (10) at (2,0)  [fill=green!40] {$G$};
\node[main node] (11) at (2,-1)  [fill=white] {$T$};
\node[main node] (12) at (2,1)  [fill=red!20] {$AF$};

\path[every node/.style={font=\sffamily\small}]
(1) edge [dashed] (2)
(1) edge [dashed] (3)
(1) edge [dashed] (4)
(2) edge node [above left] {m} (5)
(2) edge node [below] {m} (6)
(3) edge node [below] {m} (5)
(3) edge node [below] {m} (7)
(4) edge node [below] {m} (6)
(4) edge node [above right] {m} (7)
(5) edge node [above] {s} (8)
(4) edge node [right] {s} (8)
(6) edge node [below] {s} (9)
(3) edge node [right] {s} (9)
(8) edge node [below] {i} (10)
(9) edge node [below] {i} (10)
(10) edge [dashed] (11)
(7) edge node [above] {s} (12)
(2) edge node [below] {m} (12)
(12) edge [bend right, looseness=1.5, dashed] (11)
;
\end{tikzpicture}&\begin{tikzpicture}[->,x=1.5cm,y=.9cm,auto,node distance=3cm,
  thick,main node/.style={font=\footnotesize}]

\node[main node] (1) at (2,4)  [fill=white] {$S$};
\node[main node] (2) at (0.75,3)  [fill=green!40] {$A$};
\node[main node] (3) at (2,3)  [fill=green!40] {$C$};
\node[main node] (4) at (3.25,3)  [fill=green!40] {$B$};
\node[main node] (5) at (0,2)  [fill=green!40] {$D$};
\node[main node] (6) at (2,2) [fill=green!40] {$E$};
\node[main node] (7) at (4,2) [fill=green!40] {$F$};
\node[main node] (8) at (3.25,1) [fill=red!20] {$BD$};
\node[main node] (9) at (0.75,1)  [fill=red!20] {$CE$};
\node[main node] (10) at (2,0)  [fill=green!40] {$G$};
\node[main node] (11) at (2,-1)  [fill=white] {$T$};
\node[main node] (12) at (2,1)  [fill=red!20] {$AF$};

\path[every node/.style={font=\sffamily\small}]
(1) edge [dashed] (2)
(1) edge [dashed] (3)
(1) edge [dashed] (4)
(2) edge node [above left] {m} (5)
(2) edge node [below] {m} (6)
(3) edge node [below] {m} (5)
(3) edge node [below] {m} (7)
(4) edge node [below] {m} (6)
(4) edge node [above right] {m} (7)
(5) edge node [above] {s} (8)
(4) edge node [right] {s} (8)
(6) edge node [below] {s} (9)
(3) edge node [right] {s} (9)
(8) edge node [below] {i} (10)
(9) edge node [below] {i} (10)
(10) edge [dashed] (11)
(7) edge node [above] {s} (12)
(2) edge node [below] {m} (12)
(12) edge node [right] {i} (10)
;
\end{tikzpicture}
\end{tabular}
\caption{The dependency lattice of Figure 13 of the corpus (left) and the dependency lattice of Figure 13 plus an additional instruction \textsf{segment A-F intersects segment B-D at point G} (right).\label{d013bis}}
\end{figure}
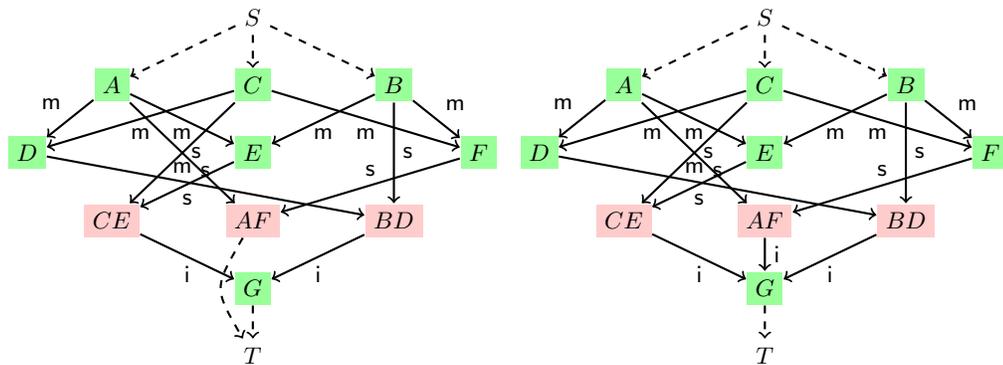

Let us now see how the dependency lattice would be affected if the XML description of Fig.~\ref{013} had an additional instruction, saying that $G$ is (also) the intersection of $BD$ and $AF$. On Fig.~\ref{d013bis} one can compare the two graphs, on the right side one can see the one with the additional instruction.

Indeed, the new dependency lattice has one additional edge $AF\to G$. On the other hand, the edge $AF\to T$ disappears since there is a path from $F$ to $T$, and $AF$ is not a sink anymore.

As dependencies have to be respected, if we remove a node from the upper part of the lattice, we will have to remove all descendants of it. For this reason, the only reasonable query reduction strategy would be to remove nodes or edges from the \emph{lower} part of the lattice.

If we remove, for example, the node $G$ (and hence the edges $CE\to G$, $AF\to G$ and $BD\to G$), then we obtain a triangle with three medians but where the barycenter has not been explicitly drawn. (Interestingly, we still obtain a figure that is visually identical to Fig.~\ref{013}.)

If we go further and remove one of the relations among $CE\to G$, $AF\to G$ and $BD\to G$ (for symmetry reasons it doesn't matter which relation we remove), then the query will succeed, while the visual representation of the figure still has not changed.

Algorithm~\ref{alg:algo} is a synthesis of the geometric figure query algorithm we propose.

\begin{algorithm}[tb]
\KwData{A corpus of declaratively described figures, a query}
\KwResult{One or more figures matching the query}

\eIf{using controlled query language}{
write the query in controlled query language\;
}{
draw the query in a DGS\;
}
convert query into ontological graph\;
apply inference to ontological graph\;
convert ontological graph to Cypher\;
submit to neo4j database\;
\If{no results returned}{
convert query into dependency lattice\;
\While{no results returned}{
extract node or relation from bottom of dependency lattice\;
remove that node or that relation from the ontological graph\;
convert ontological graph to Cypher\;
submit to neo4j database\;
}
}

\medskip

\caption{The Query Algorithm for a Declaratively Described Corpus.}\label{alg:algo}
\end{algorithm}

\section{Evaluation}

%[At the time of submission of the article, evaluation was not completed, hence it is presented as a plan.]}

We plan to evaluate the algorithms described in this paper, in the following ways:

Querying a sub-figure in a corpus of declaratively described figures gives a binary result: either the figure is matching the sub-figure or it is not, so evaluation is simply counting the number of successes.

An interesting parameter to observe is the number and nature of query reductions that were necessary to obtain results, correlated with the number of results obtained.

We will proceed as follows: after visually inspecting the corpus (and hence with no knowledge about the procedural and declarative descriptions of the figures) we will formulate 20 queries and manually annotate the figures we expect to find.

After using the algorithm, we will count the number of successes and study the number of results vs. the parameters of query reduction.

\section{Future Work}
As future work, besides extending the controlled natural language (\S\,\ref{fpcnl}), we plan to integrate this search mechanism in repositories such as TGTP and Intergeo, and in learning environments like the Web Geometry Laboratory~\cite{Quaresma2013}. In a more generic approach, we should use a common format and develop an application programming interface that will allow to integrate the searching mechanism in any geometric system in need of it. 
 
\section{Conclusion}
In this paper we have presented algorithms for querying geometric figures in either declaratively or analytically described corpora, by using either a DGS or a dedicated controlled query language.

At the time of submission of the article, evaluation was not completed, hence it is presented as a plan.

\bibliographystyle{splncs03}
\bibliography{cicm2014}

\begin{thebibliography}{10}
\providecommand{\url}[1]{\texttt{#1}}
\providecommand{\urlprefix}{URL }

\bibitem{PLY}
Beazley, D.: Python {Lex-Yacc}, \url{http://www.dabeaz.com/ply/}

\bibitem{Chou2001}
Chou, S.C., Gao, X.S.: Automated reasoning in geometry. In: Handbook of
  Automated Reasoning. pp. 707--749. Elsevier Science Publishers (2001)

\bibitem{Chou2000}
Chou, S.C., Gao, X.S., Zhang, J.Z.: A deductive database approach to automated
  geometry theorem proving and discovering. Journal of Automated Reasoning  25,
   219–246 (2000)

\bibitem{diffeor}
DiffeoR: Answer to ``{I}s it possible to reconstruct a triangle from the
  midpoints of its sides?'' (2014/2/20),
  \url{http://math.stackexchange.com/a/683496/122762}

\bibitem{Fagin:1983dy}
Fagin, R.: Degrees of acyclicity for hypergraphs and relational database
  schemes. Journal of the ACM (JACM)  30(3),  514--550 (Jul 1983)

\bibitem{geogebra}
Hohenwarter, M., Preiner, J.: Dynamic mathematics with {GeoGebra}. The Journal
  of Online Mathematics and Its Applications  7 (2007), {ID 1448}

\bibitem{Janicic06a}
Jani\v{c}i\'c, P., Quaresma, P.: System description: {GCLC}prover +
  {G}eo{T}hms. In: Furbach, U., Shankar, N. (eds.) Automated Reasoning. Lecture
  Notes in Computer Science, vol. 4130, pp. 145--150. Springer (2006)

\bibitem{Jiang2012}
Jiang, J., Zhang, J.: A review and prospect of readable machine proofs for
  geometry theorems. Journal of Systems Science and Complexity  25,  802--820
  (2012)

\bibitem{Kortenkamp2009}
Kortenkamp, U., Dohrmann, C., Kreis, Y., Dording, C., Libbrecht, P., Mercat,
  C.: Using the {I}ntergeo platform for teaching and research. In: Proceedings
  of the 9th International Conference on Technology in Mathematics Teaching
  (ICTMT-9) (2009)

\bibitem{Quaresma08}
Quaresma, P., Jani\v{c}i\'c, P., T.J., Vujo\v{s}evi\'c-Jani\v{c}i\'c, M.,
  To\v{s}i\'c, D.: Communicating Mathematics in The Digital Era, chap.
  XML-Bases Format for Descriptions of Geometric Constructions and Proofs, pp.
  183--197. A. K. Peters, Ltd. (2008)

\bibitem{Quaresma2011}
Quaresma, P.: {T}housands of {G}eometric problems for geometric {T}heorem
  {P}rovers ({TGTP}). In: Schreck, P., Narboux, J., Richter-Gebert, J. (eds.)
  Automated Deduction in Geometry, Lecture Notes in Computer Science, vol.
  6877, pp. 169--181. Springer (2011)

\bibitem{Quaresma2012}
Quaresma, P.: An {XML}-format for conjectures in geometry. pp. 54--65. No. 921
  in CEUR Workshop Proceedings, Aachen (2012),
  \url{http://ceur-ws.org/Vol-921/}

\bibitem{H-Q2012}
Quaresma, P., Haralambous, Y.: Geometry construction recognition by the use of
  semantic graphs. In: RECPAD 2012: 18th Portuguese Conference on Pattern
  Recognition, Coimbra: 26-26 October 2012. pp. 47--48 (2012)

\bibitem{Quaresma2013}
Quaresma, P., Santos, V., Bouallegue, S.: The {W}eb {G}eometry laboratory
  project. In: CICM 2013 Proceedings. Lecture Notes in Artificial Intelligence,
  vol. 7961, pp. 364--368. Springer (2013)

\bibitem{Cypher}
Robinson, I., Webber, J., Eifrem, E.: Graph Databases. O'Reilly (2013)

\bibitem{Santiago2010}
Santiago, E., Hendriks, M., Kreis, Y., Kortenkamp, U., Marqu\`es, D.: {\sc i2g}
  {C}ommon {F}ile {F}ormat {F}inal {V}ersion. Tech. Rep. D3.10, The Intergeo
  Consortium (2010), \url{http://i2geo.net/xwiki/bin/view/I2GFormat/}

\bibitem{i2gImplementationTable}
{The Intergeo Consortium}: Intergeo implementation table.
  \url{http://i2geo.net/xwiki/bin/view/I2GFormat/ImplementationsTable} (2012)

\bibitem{wikipediaLIGS}
Wikipedia: List of interactive geometry software.
  \url{http://en.wikipedia.org/wiki/List_of_interactive_geometry_software}
  (February 2014)

\end{thebibliography}

\end{document}